\documentclass[final]{aipproc}
\layoutstyle{6x9}

\begin{document}

\title{Afterglows of Gamma-Ray Bursts: Short vs. Long GRBs}


\author{David Alexander Kann \& Sylvio Klose (on behalf of a larger collaboration)} 
       {address={Th\"uringer Landessternwarte Tautenburg, 07778
        Tautenburg, Germany}}

\begin{abstract}
We compiled a large sample of \emph{Swift}-era photometric data on long (Type II) and short (Type I) GRB afterglows. We compare the luminosity and energetics of the different samples to each other and to the afterglows of the pre-\emph{Swift} era. Here, we present the first results of these studies.
\end{abstract}

\maketitle

\section{Introduction}

In a systematic analysis of the afterglows of (Type II) GRBs up to the launch of \emph{Swift}, Kann et al. 2006, Liang \& Zhang 2006 and Nardini et al. 2006 reported that afterglows cluster at fixed (late, e.g., 0.5 rest-frame days) times in luminosity space. Furthermore, there is a bimodality, i.e., two tight clusters, that are clearly separated in terms of redshifts, with most less luminous afterglows having $z\leq1.4$ and most afterglows of high luminosity having $z\ge1.4$. We have expanded the data collection presented in Kann et al. 2006 to afterglows of \emph{Swift}-era Type II GRBs (also detected by satellite missions other than \emph{Swift}), and to the afterglows of Type I GRBs (which are usually attributed to the merger of compact objects), which were first discovered in the \emph{Swift} era. The in-depth analysis and results are presented in Kann et al. 2007, which deals with the comparison between pre-\emph{Swift} and \emph{Swift}-era Type II GRB afterglows, and in Kann et al., 2008, where the afterglows of Type II (the complete sample) and Type I GRBs are compared.

\section{A comparison between pre-\emph{Swift} and \emph{Swift}-era Type II GRB afterglows}

Following the methods presented in Kann et al. 2006, we analyzed the light curves of a total of 38 Type II GRB afterglows. For 14 of these, we were able to derive spectral energy distributions in the optical range (and sometimes into the UV and/or NIR) that fulfilled the criteria of the ``Golden Sample'' of Kann et al. 2006. Fitting the SEDs with dust extinction models we were able to determine the host galaxy extinction along the line of sight. For another 9 afterglows, we were able to determine rough extinction values, and for the final 15, we only had light curve data and assumed no extinction. All GRBs have measured redshifts. Knowledge of the redshift and the line-of-sight extinction allowed us to shift all afterglow light curves to a common redshift $z=1$ and directly compare their luminosities.

\begin{figure}
\includegraphics[height=.68\textheight]{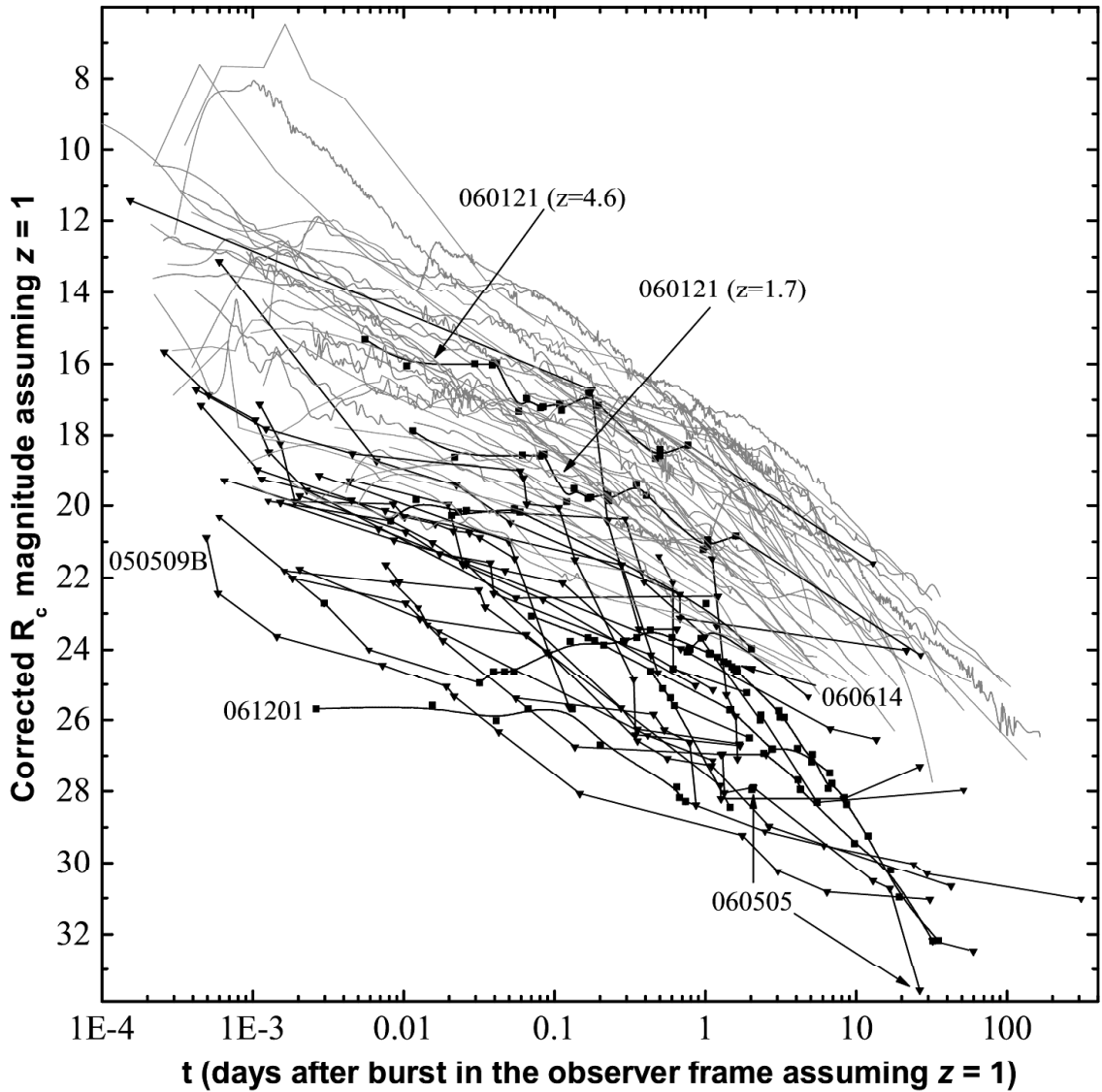}
\caption{The afterglows of Type II (gray) and Type I (black) GRBs. All afterglows have been corrected for extinction (where measurable) and shifted to a common redshift $z=1$, so that their luminosities and evolution are directly comparable.}
\end{figure}

We found (see Kann et al. 2007 for more details) that intrinsically, the afterglows of the pre-\emph{Swift} and \emph{Swift}-era are very similar to each other. In most cases, the best-fitting dust model resembled Small Magellanic Cloud dust, and the line-of-sight extinction was typically low ($A_V\approx0.2$). Once again, there was a clustering of luminosities visible, and once again, nearby afterglows were less luminous than more distant ones. The fact that \emph{Swift}-era GRB afterglows are observationally fainter than those of the pre-\emph{Swift} era is mostly an effect of the higher mean redshift (Jakobsson et al 2006). We were able to study the luminosities at very early times (minutes after the GRB), which was not possible in the pre-\emph{Swift} era, and found a broad spread of luminosities. On the other hand, over half of all afterglows clustered tightly even at such early times. We found indications of an exponential cutoff of the luminosity distribution at high luminosities. The afterglows of Type II GRBs are shown in Fig. 1.

\section{A comparison between Type II and Type I GRB afterglows}

We also compiled all available photometry on all Type I GRB afterglows until November 2007, both detections and upper limits. In many cases, no redshift was known, and in almost all cases, we had no information on extinction in the hosts. Therefore, we had to make assumptions (see Kann et al. 2008 for more details). The afterglows were shifted to $z=1$ and compared to the afterglows of Type II GRBs.

Fig. 1 also shows the afterglows (or upper limits in case of non-detections) of Type I GRB afterglows. Clearly, almost all of them are much less luminous than the afterglows of Type II GRBs. A clustering is not visible, the spread is even increased in comparison to the observed distribution. We derived the magnitudes at one day after the GRB (assuming $z=1$) and transformed them to absolute magnitudes and found that, in the mean, the afterglows of Type I GRBs are five magnitudes, a factor of 100 in flux density, fainter than those of Type II GRBs. This is visible in the magnitude distribution shown in Fig. 2. The sparse detections did not allow us to determine if there are fundamental differences in light curve evolution between the afterglows of the different progenitors.

\begin{figure}
\includegraphics[height=.3\textheight]{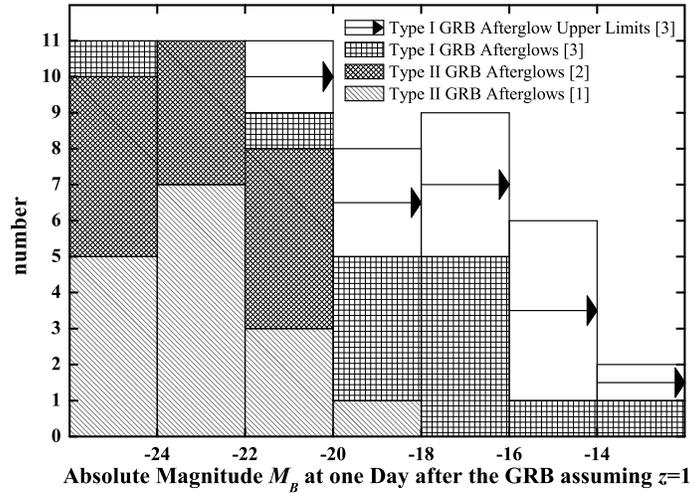}
\caption{The absolute magnitude $M_B$ of all afterglows, measured at 1 day (assuming $z=1$) after the GRB. [1], [2], [3]: Kann et al. 2006, 2007, 2008, respectively.}
\end{figure}

We also compiled data on the high energy prompt emission of the GRBs in all our samples. We performed a bolometric correction of the isotropic energy release. In Fig. 3, we show the flux density, again at one day assuming $z=1$, plotted against the isotropic energy release. Type II GRBs populate the top right of the plot, and Type I GRBs the lower left, with only a few cases of overlap. These are either energetic Type I GRBs, or subenergetic Type II GRBs. Using all detected afterglows (where we only chose those Type I GRB afterglows with a secure redshift, but note that those without secure redshift are not strong outliers), a correlation is visible which becomes steeper if the sample consists of both Type I and Type II GRB afterglows. Adding upper limits on Type I GRB afterglows with secure redshifts makes the correlation steeper still.

\begin{figure}
\includegraphics[height=.3\textheight]{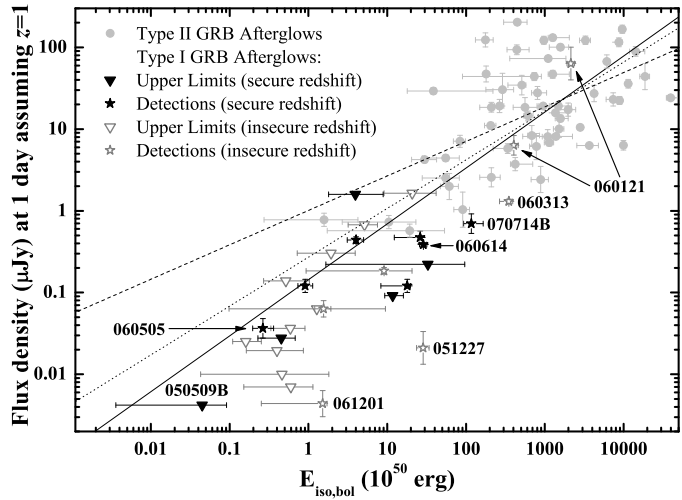}
\caption{The optical flux density (measured at 1 day, assuming $z=1$, after the GRB) versus the bolometric isotropic energy release of the prompt emission. A correlation with a lot of scatter is visible. Dashed: Type II only. Dotted: also the Type I detections. Straight: also the Type I upper limits (in both cases only with secure redshifts).}
\end{figure}

In Kann et al. 2008, we also present a systematic study of (the absence of) supernova light in Type I GRB afterglows, both from classical broad-lined Type Ic SNe (associated with Type II GRBs) and from ``mini-supernovae'' as proposed by Li \& Paczy\'nski 1998. Furthermore, we discuss several controversial GRBs in the light of their afterglow luminosities, and look for further correlations between the afterglow luminosities, the energetics and other GRB parameters such as duration and host galaxy offset.


\begin{theacknowledgments}
D.A.K. and S.K. acknowledge financial support from DFG Kl 766/13-2. 
\end{theacknowledgments}


\end{document}